# Ambiente de Simulações com Traçado de Raios para mmWave MIMO Aplicado a VANTs

Felipe Henrique Bastos e Bastos, Ingrid Nascimento e Aldebaro Klautau

*Resumo*— O uso de VANTs tende a aumentar nos próximos anos, assim é de grande importância estudar formas de comunicações com estes aeromodelos. Esse trabalho propõe um ambiente de simulações que combina simuladores de tráfego, voo e traçado de raios para a geração de grandes quantidades de dados de propagação entre transmissores terrestres e veículos aéreos em redes 5G.

*Palavras-Chave*— MIMO, Ondas milimétricas, VANT, Simulação de sistemas de comunicação.

*Abstract*— Some applications and use cases of unmanned aerial vehicle (UAV or drones) in 5G require communication channels with large capacities to transmit, for example, high-resolution videos in real-time. In this case, millimeter waves (mmWaves) are important given the large bandwidth and, consequently, potential to deliver very high bit rates especially when used together with multiple antenna MIMO systems. In order to enable UAVs communicating via mmWaves MIMO, it is important to have good channel models and radio propagation data to guide the system design. This paper presents a simulation environment that combines simulators of traffic, flight and ray tracing, to generate large amounts of propagation data between terrestrial transmitters and aerial vehicles in 5G networks. Preliminary results show that the proposed environment is flexible and allow for several specific investigations.

*Keywords*— MIMO, mmWave, UAV, Simulation of communication systems.

## I. Introdução

O uso de Veículos Aéreos Não Tripulados (VANT) tende a aumentar em diversas áreas, tal como monitoramento de plantações, entrega de objetos, segurança de patrimônio, etc [2]. Além disso, estes podem ser usados em redes 5G como retransmissores, provedores locais de conteúdo, dentre outras aplicações [3].

O principal objetivo deste trabalho é desenvolver um ambiente de simulações para facilitar estudos de VANT que usem comunicações baseadas em ondas milimétricas e sistemas 5G com Múltiplas Entradas e Múltiplas Saídas (MIMO). Para isso, está sendo desenvolvido com base na metodologia proposta em [1], um software orquestrador que controla simulações integrando o simulador Wireless InSite (WI) da empresa Remcom, que utiliza traçado de raios, com alguns simuladores de vôo como o AirSim. Ao fim de uma simulação, o resultado final é um banco de dados contendo diversas informações dos canais a partir dos raios propagados durante a simulação.

## II. Simulações

### A. Ambiente de Simulção

Inicialmente usa-se o "Simulation of Urban MObility" (SUMO), um simulador de trafico de veículos e pedestres para gerar a posição dos VANTs no plano. Em seguida os VANTs são desenhados no WI em uma altura definida nas configurações da simulação. A simulação no WI é executada e os dados de saída gravadas em um banco de dados SQL. A Fig. 1 ilustra o processo.

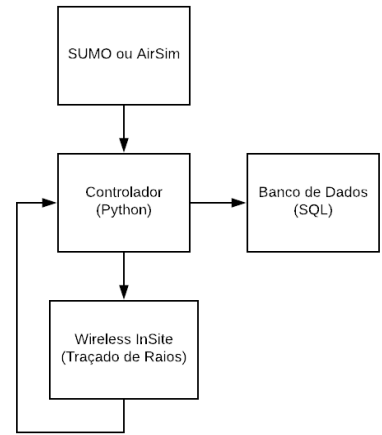

Fig. 1
DIAGRAMA DO FLUXO DAS SIMULAÇÕES.

### B. Estudo de caso: Cenário urbano

Para simulações iniciais foi utilizado um cenário urbano, o mesmo usado em [1]. Aqui porém os novos módulos do software desenvolvidos permitiram que carros pudessem ser substituídos por VANTs.

Foram adicionadas rotas para os VANTs no SUMO que pode ser vistas na Figura 2 . Foram executadas simulações inserindo/adicionando os VANTs em três diferentes alturas: 50m, 100m e 150m. A altura dos aeromodelos é somada no momento de desenhá-los no WI.

Os prédios são todos de concreto e os aeromodelos de metal. Todas as antenas receptoras foram são posicionadas abaixo do VANTs e a antena transmissora na calçada a 5m do chão.

A área de estudos definida possui duas ruas, 20 prédios com alturas de 10m a 90m e 10 VANTs, divididos aleatoriamente em 3 modelos diferentes.

Na Tabela 1 é possível encontrar em resumo com os principais parâmetros utilizados nas simulações.

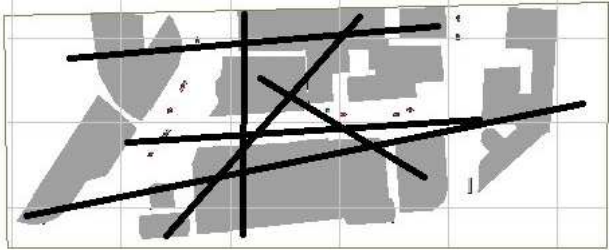

Fig. 2
ROTAS DOS VANTs.

TABELA I
PARAMETROS DE SIMULAÇÃO.

| Parametros WI | |
|---|---|
| Frequência da portadora | 60 GHz |
| Potência transmitida | 0 dBm |
| Altura da antena transmissora | 5 m |
| Antenas (Tx e Rx) | Dipolo de meia onda |
| Modelo de propagação | X3D |
| Material dos prédios | ITU concreto 60 GHz |
| Material dos VATs | Metal |
| Espaçamento entre raios (graus) | 1 |
| Num. $L$ de raios analisados | 25 |
| Modelo de diffuse scattering | Lambertian |
| Reflexão max. De DS ($N_{max}^{DS}$) | 2 |
| Coeficientes para o DS ($S$) | 0.4 (concreto), 0.2 (metal) |
| Parametros VANTs | |
| Número de rotas | 6 |
| VANTs | entregas, domestico, rural |
| Comprimetos, respectivamente (mm) | 914, 289.5, 716 |
| Larguras, respectivamente (mm) | 914, 289.5, 220 |
| Alturas, respectivamente (mm) | 336, 196, 236 |
| Velocidade max., respectivamente (m/s) | 24.5, 20.0, 23.0 |
| Paiodo de amostragem no SUMO $T_{sam}$ (s) | 0.1 |

## III. RESULTADOS

### A. Banco de Dados

A partir do ambiente de simulação montado foi possível obter um grande Banco de Dados (BD) com diversas informações sobre os raios propagados da antena transmissora até as receptoras, como: potência recebida, ângulos de chagada e saída do sinal, eventos de propagação como reflexões, difrações, etc.

No BD as informações ficam divididas em episódios, cenas, receptores e raios, como detalhado em [1]. Associado a cada raio é possível visualizar os dados anteriormente citados. Para ilustrar, as Figs. 4 e 5 mostram gráficos da potência recebida e do atraso de propagação para um VANT específico.

## IV. CONCLUSÃO

O ambiente de simulações desenvolvido é capaz de gerar grandes quantidades de dados de propagação em cenários envolvendo aeromodelos e carros. Os cenários-base, com os prédios, ruas, etc., são obtidos de sites especializados em CAD na Web. Os modelos de VANTs e carros podem ser facilmente alterados e modelados para as mais diversas simulações. Os dados gerados podem ser usados para elaboração de modelos de canais de propagação visando novas aplicações em redes 5G.

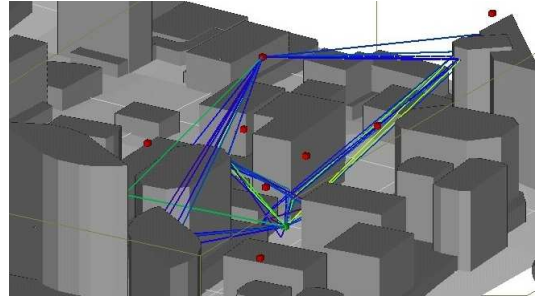

Fig. 3
RAIOS RECEBIDOS POR UM VANT APÓS SIMULAÇÃO.

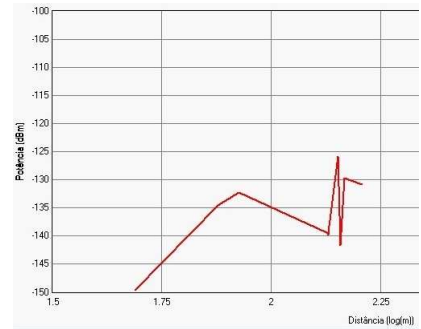

Fig. 4
GRÁFICO DA POTÊNCIA RECEBIDA POR UM VANT A 100M.

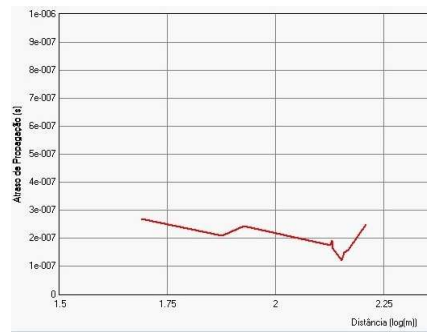

Fig. 5
GRÁFICO DO ATRASO DE PROPAGAÇÃO DO SINAL RECEBIDO POR UM VANT A 100M.